\documentclass[aps,prA,amsmath,amsfonts,10pt,twocolumn]{revtex4}
\usepackage{epsfig,graphicx}
\usepackage[english]{babel}
\usepackage{amsfonts}
\usepackage{amsmath}
\usepackage{latexsym}
\usepackage{graphics,bm}
\usepackage{subfigure}
\usepackage{dcolumn}
\usepackage{bm}
\usepackage{rotating}

\newcommand{\vnu}{V_{{0}}}
\newcommand{\nnu}{n_{0}}
\newcommand{\hsi}{\hbar\Sigma_{12}}
\def\bp{{\bf p}}

\def\bx{{\bf x}}

\def\bk{{\bf k}}

\def\rT{{\rm T}}
\def\bu{{\bf u}}
\def\rTtb{{\rm T}^{2B}}
\def\rTmb{{\rm T}^{MB}}
\def\rTmbu{{\rm T}}

\def\Rk{R_0}
\def\velu{{\bf u}}
\def\velv{{\bf v}_{s}}
\def\veln{{\bf v}_{n}}
\def\muef{\mu_{\rm eff}}
\def\vk{v_{\bf k}}
\def\uk{u_{\bf k}}

\newcommand{\ndis}{ n_{\rm R}}

\newcommand{\mur}{ \delta\mu_{\rm R}}
\def\epsk{\epsilon_{\bf k}}
\def\Ek{\hbar\Omega_{{\bf k}}}
\def\Emenok{\hbar\Omega_{-{\bf k}}}
\def\ddt{\frac{\partial}{\partial t}}
\def\tcz{T_c^{0}}

\def\at{a_{\rm T}}
\begin{document}

\title{Thermodynamics of a Bose-Einstein Condensate with Weak Disorder}

\author{G. M. Falco, A. Pelster, R. Graham}
\affiliation{Universit\"at Duisburg-Essen \\ Fachbereich Physik, Campus Duisburg
\\Lotharstrasse 1, 47057 Duisburg, Germany }

\begin{abstract}
We consider the thermodynamics of a homogeneous superfluid 
dilute Bose gas in the presence
of weak quenched disorder.
Following the zero-temperature approach of Huang and Meng, 
we diago\-nalize
the Hamiltonian of a dilute 
Bose gas in an external random
delta-correlated potential by means of a Bogoliubov transformation.  
We extend this approach to finite temperature 
by combining
the Popov and  the many-body ${\rm T}-$matrix approximations.
This approach permits us to include the quasi-particle interactions  
within this temperature range. We derive the disorder-induced shifts of 
the Bose-Einstein critical
temperature and of the temperature for the onset of superfluidity 
by approaching the transition points from below, i.e., from the superfluid phase.
Our results lead to a phase diagram 
consistent
with that of the finite-temperature theory of Lopatin and Vinokur
which was
based on the replica method, and in which the transition points were approached from above. 
\end{abstract}

\date{\today}



\maketitle

\section{Introduction}
An interacting ultracold dilute Bose gas in a weak random external potential,
which is homogeneous in the mean,
represents an interesting model for studying the relation between Bose-Einstein
condensation and superfluidity and
has been the subject of various theoretical
investigations in the last few years \cite{Huang92,Giorgini93,Lopatin02,Kobayashi02,Pelster05,Zobay06}.
Two different methods have been used
for performing
the average over the impurity scatterers.
In Refs.~\cite{Huang92,Giorgini93,Kobayashi02} the avera\-ge of the grand potential over the
disorder is taken perturbatively in the strength of the disorder
after diagonalizing the Hamiltonian by means 
of a Bogoliubov transformation.
Alternatively, in Refs.~\cite{Lopatin02,Pelster05,Zobay06} the avera\-ging is implemented 
by the replica method. 
In Ref.~\cite{Lopatin02}, the replica symmetric solution of the model
is found by a systematic diagrammatic Beliaev-Popov perturbation theory
for the dilute superfluid gas in the presence of disorder.
At zero temperature the two different approaches 
give equivalent results.
At finite temperatures, however, the approach based on the Bogoliubov transformation
becomes unsa\-tisfactory since it
necessarily neglects important quasi-particles correlations 
\cite{Huang92,Kobayashi02}.
The replica trick, on the other hand, has  limitations of it's own by making
 the mathematical and also the physical description less transparent.
For these reasons it would clearly be desirable to develop an alternative theory
avoiding these shortcomings. 
This is the goal of this paper, where we
show how the perturbative approach
of Refs.~\cite{Huang92,Giorgini93,Kobayashi02} can be 
extended
to include the leading effects of the scattering between quasi-particles 
at finite temperature. As a result, some discrepancies 
between the two different methods are resolved. 

The paper is organized as follows.
In order to make the paper self-contained,
we briefly rederive in Section~\ref{sec:Bog} 
the thermodynamic 
potential and the equation of state for the case of a vanishing spatial correlation length
of the disorder potential \cite{Huang92}.
In the limit of zero temperature, we also determine the high-order Beliaev corrections
to the chemical potential \cite{Lopatin02}.
In Section~\ref{sec:superfluidity} we give the derivation of the superfluid component 
of the system
in the presence of a random potential. 
The disorder-induced corrections of the velocity of sound are there
obtained from
hydrodynamic equations for the superfluid component of the system. 
In Section~\ref{sec:popov}
we extend the theory 
to finite temperatures within the mean-field Popov approximation. 
In Section~\ref{sec:many-bodyTmatrix} the thermodynamics and the phase 
diagram 
are investigated 
%
by means of the many-body $\rTmbu-$matrix approximation.
In particular, we use the theory to calculate the shift 
of the Bose-Einstein critical temperature  
and of the temperature for the loss of the superfluidity when
approaching the critical points from below.
Comments and conclusions remain for Section~\ref{sec:conc}. 

\section{Bogoliubov's theory}
\label{sec:Bog}

We consider the effects of an external random field on the thermodynamics of a dilute 
Bose gas. The random field is assumed to have a 
probability distribution $P[ U]$
normalized to one
when averaged over all disorder configurations, 
that is $\int d\left[ U\right]P[ U]=1$.
The average over the disorder fields is defined as
\begin{align}
\label{eq:disaverage0}
\langle \bullet\rangle=\int d\left[ U\right]\bullet P[ U],
\end{align}
and for the disorder potential we assume the ensemble averages
\begin{align}
\label{eq:noise1}
\langle U(\bf x)\rangle&=0,\nonumber\\
\langle U({\bf x})U({\bf x}')\rangle&=R\left({\bf x}-{\bf x'}\right).
\end{align}
We assume that any time scale  of the disorder potential
is frozen, i.e. very long in comparison with the 
thermodynamic time scale.
This so-called quenched-disorder limit 
has the consequence that the disorder average must be taken after the 
thermodynamic average over the grand-canonical ensemble. Therefore, 
the total average of an observable
reads
\begin{align}
\label{eq:disaverage1}
\langle \langle O\rangle_{\rm gr}\rangle=\int d\left[ U\right] P[ U]\langle O(\Psi)\rangle_{\rm gr},
\end{align}
where $\langle O(\Psi)\rangle_{\rm gr}$ indicates the grand-canonical average.
At very low temperatures, when the wavelength of the atoms in the gas is much larger than 
the range of the impurity scatterers responsible for the random potential \cite{Astra01},
one may consider the limit of a $\delta-$correlated type of disorder 
$R\left({\bf x}-{\bf x'}
\right)=R_0~\delta({\bf x}-{\bf x}')$.
The constant $R_0$ is then related to the concentration of the impurities and to the 
$s-$wave scattering length of the random scatterers \cite{Astra01}.

At equilibrium, 
the grand-canonical partition function of
a Bose gas in a disordered medium is given as a functional of the
random external potential $ U(\bx)$  
\begin{equation}
\label{eq:partfunct}
Z_{\rm gr}[U]
=\int
d[\psi^*]d[\psi]\exp\left\{-\frac{1}{\hbar}S[\psi^*,\psi;U]\right\}~,
\end{equation}
where the functional integral is performed over $c$-number fields
$\psi^*({\bf x},\tau)$ and $\psi({\bf x},\tau)$ periodic in
imaginary time over $\hbar\beta=\hbar/k_BT$.
The Euclidean action $S$ is given by 
\begin{align}
\label{eq:action}
S[\psi^*,\psi;U]&=\\ \int_{0}^{\hbar\beta}d\tau \!\!&\int d{\bf x}~\psi^*({\bf x},\tau)
\left\{\left[ \hbar\frac{\partial}{\partial\tau}-
\frac{\hbar^2{\bf \nabla}^2}{2m}-\mu +U(\bx)\right] \right.
\nonumber \\ 
+&\left.\frac{1}{2}\int d{\bf x'}~\psi^*({\bf
x'},\tau)V({\bf x}-{\bf x'})\psi({\bf x'},\tau) \right\}\psi({\bf x},\tau)~\nonumber,
\end{align}
where $\mu$ is the chemical potential 
and $V({\bf x}-{\bf x'})$ the
atomic repulsive interaction potential.
The disorder average of the thermodynamic potential 
$\Omega=-\left({\rm ln} Z_{\rm gr}\right)/\beta$
is obtained from 
\begin{equation}
\label{eq:averageOmega}
\langle\Omega\rangle=-\frac{1}{\beta}\langle{\rm ln}Z_{\rm gr}\rangle.
\end{equation}
The average of Eq.~(\ref{eq:averageOmega})
is highly non-trivial because the disorder average is nonlinear in $Z_{\rm gr}$
due to the logarithm.
In this paper we calculate this average by following the method of 
Huang and Meng in Ref.~\cite{Huang92} which is
based on a 
canonical Bogoliubov transformation \cite{Bogoliubov47}.   

Expanding the fields in Fourier modes in a finite volume $V$ according to
$\psi({\bf x},\tau)=\left(1/\hbar\beta V\right)^{1/2}\sum_{{\bf k},n} a_{{\bf
k},n}e^{i({\bf k\cdot x}-\omega_n\tau)}$
with the Matsubara frequencies 
$\omega_n=2\pi n/\hbar\beta$,
and the corresponding complex conjugate expression for $\psi^*({\bf x},\tau)$, we 
write the action in momentum space as
\begin{align}
\label{eq:actionmomspace}
S[a^*,a]=& 
\sum_{{\bf k},n}(-i\hbar\omega_n+\epsilon_{{\bf k}}-\mu)
a^*_{{\bf k},n}a_{{\bf k},n}\\
&+
\frac{1}{V}\sum_{{\bf
k,k'},n}a^*_{{\bf k},n}U_{{\bf k}-{\bf k}'}
a_{{\bf k}',n}\delta_{n,0}
\nonumber\\
+\frac{1}{2\hbar\beta V}&
\sum_{\stackrel{\bf
k,k',q}{n,n',m}} V_{\bf q}~a^*_{{\bf k+q},n+m}a^*_{{\bf k'-q},n'-m}
a_{{\bf k'},n'}a_{{\bf k},n}~.\nonumber
\end{align}
In this equation we have introduced 
the free particle dispersion $\epsilon_{{\bf k}}=
\hbar^2{\bf k}^2/2m$ and the Fourier transform of the
interaction potential $V_{\bf q}=\int d{\bf x}~V({\bf x})e^{-i{\bf q\cdot
x}}$. At very low temperatures,
the de Broglie wavelength $\lambda_{\rm th}$ of the atoms
is much larger than the range $r_0$ of the interaction potential
such that $r_0/\lambda_{\rm th}\ll 1$.
Therefore, only $s$-wave scattering is relevant
in the gas and we
can neglect the momentum dependence of the interaction potential,
setting $V_{\bf q}=V_0$. 
Breaking
the gauge-symmetry of the action by introducing the decomposition 
$\psi({\bf x},\tau)=\sqrt{n_0}+
\psi'({\bf x},\tau)$
and expanding the resulting expression  up to
quadratic order in $\psi'$ and $U$, the effective action becomes
\begin{align}
\label{eq:quadract}
S^{(2)}[a,a^*]&=-\hbar\beta\mu n_0 V+\frac{1}{2}\hbar\beta n_0^2 V_0 V+
\hbar\beta{n_0}
U_{{\bf 0}}\\
&+
{{\sum_{{\bf k},n}}}^{'}(-i\hbar\omega_n+\epsilon_{{\bf
k}}-\mu+2n_0 V_0)a^*_{{\bf k},n}a_{{\bf k},n}\nonumber \\
&+\left(
\frac{n_0}{V}\right)^{\frac{1}{2}}{\sum_{{\bf
k},n}}^{'}\left(a_{{\bf k},n}^*U_{{\bf k}}
+U_{-{\bf k}}a_{{\bf k},n}\right)\delta_{n,0}\nonumber\\
&+\frac{1}{2}n_0 V_0{\sum_{{\bf k},n}}^{'}(a^*_{{\bf k},n}a^*_{{\bf -k},-n}
+a_{{\bf k},n}a_{{\bf -k},-n})~,\nonumber
\end{align}
where the prime denotes that ${\bf k}={\bf 0}$ is excluded from the sum.
The condensate density $n_0$ remains to be determined by 
minimizing the thermodynamic potential.
Note that we have also neglected as of higher order the
term $1/V {\sum^{'}_{{\bf k},{\bf k}',n}} a^*_{{\bf k},n} ~ U_{{\bf k}-{\bf k}' }~ a_{{\bf k}',n}\delta_{n,0}$,
which requires the assumption of weak disorder \cite{Huang92}.
The effective action is diagonalized by a Bogoliubov transformation 
\begin{align}
\label{eq:Bogtrans}
a_{{\bf k},n}=u_{{\bf k}}\alpha_{{\bf k},n}-v_{{\bf k}}\alpha_{{-\bf k},-n}^*-z_{{\bf k}}\nonumber\\
a_{{\bf k},n}^*=u_{{\bf k}}^*\alpha_{{\bf k},n}^*-v_{{\bf k}}^*\alpha_{{-\bf k},-n}-z_{{-\bf k}}^*,
\end{align}
where the coherence factors $u_{{\bf k}}$, $v_{{\bf k}}$ 
and the complex number $z_{{\bf k}}$ can be taken real positive by appropriately choosing 
the phase of the complex fields.
In such a case we have 
\begin{align}
\label{eq:uandv}
u_{{\bf k}}^2=\frac{1}{2}\left[1+\frac{\epsilon_{\bf k}-\mu+2n_0V_0
}{\hbar\Omega_{\bf k}}\right]\nonumber\\
v_{{\bf k}}^2=\frac{1}{2}\left[-1+\frac{\epsilon_{\bf k}-\mu+2n_0V_0
}{\hbar\Omega_{\bf k}}
\right]\nonumber\\
z_{\pm{\bf k}}=\left(n_0/V\right)^{\frac{1}{2}}\frac{U_{\bf k}}{\hbar\Omega_{\bf k}}\left(
u_{{\bf k}}-v_{{\bf k}}\right)^2,
\end{align}
where
$|u_{{\bf k}}|^2-|v_{{\bf k}}|^2=1$, $u_{{\bf k}}v_{{\bf k}}\geq 0$,
and where the Bogoliu\-bov spectrum is given by 
\begin{align}
\label{eq:bogospect}
\hbar\Omega_{\bf k}=\sqrt{\left(\epsilon_{\bf k}-\mu+2n_0V_0
\right)^2-\left(n_0 V_0\right)^2}.
\end{align}
After the diagonalization the action reads
\begin{align}
S^{(2)}&[a,a^*]
=-\hbar\beta\mu n_0 V+\frac{1}{2}\hbar\beta n_0^2 V_0 V+\hbar\beta n_0 U_{{\bf 0}}\nonumber\\
&+
{{\sum_{{\bf k},n}}}^{'}\alpha_{{\bf k},n}^*\alpha_{{\bf k},n}~\left(
-i\hbar\omega_n+\hbar\Omega_{\bf k}\right)
\nonumber\\
&+\frac{\beta\hbar}{2}{{\sum_{{\bf k}}}}^{'}\left[\Omega_{\bf k}-\left(\epsilon_{{\bf
k}}-\mu+2n_0 V_0
\right)\right]\nonumber\\&-\frac{\beta\hbar}{V}
{{\sum_{{\bf k},n}}}^{'}|U_{\bf k}|^2
\frac{n_0\left(\epsilon_{{\bf
k}}-\mu+n_0 V_0
\right)}{\hbar\Omega_{\bf k}^2}.
\end{align}
By performing the functional integral of Eq.~(\ref{eq:partfunct})
and averaging over the disorder, the thermodynamic potential 
of Eq.~(\ref{eq:averageOmega}) becomes
\begin{align}
\label{eq:thermpotBog0}
\langle\Omega\rangle
=&-\mu n_0 V+\frac{1}{2}n_0^2 V_0 V+
{{\sum_{{\bf k}}}}^{'}\frac{1}{\beta}\rm{ln}\left(1-{\it e}^{-\beta\hbar\Omega_{{\bf
k}}
}
\right)
\nonumber\\
&+\frac{1}{2}{{\sum_{{\bf k}}}}^{'}\left[\hbar\Omega_{\bf k}-\left(\epsilon_{{\bf
k}}-\mu+2n_0 V_0
\right)\right]\nonumber\\&-
{{\sum_{{\bf k}}}}^{'}n_0 \Rk
\frac{\left(\epsilon_{{\bf
k}}-\mu+n_0 V_0
\right)}{\hbar\Omega_{\bf k}^2}.
\end{align}
The first two terms on the r.h.s. represent the mean-field result while the remaining three terms
describe thermal and quantum fluctuations.
From the thermodynamic relation $n=-\left(1/V\right)\partial\langle\Omega\rangle/\partial \mu$
we have for 
the particle-density in the grand-canonical ensemble
\begin{align}
\label{eq:parnumBog}
n= n_0+n'+\ndis, 
\end{align} 
with the depletion due to the normal interaction
\begin{align}
\label{eq:nprime}
n'=&
\frac{1}{V}{{\sum_{{\bf k}}}}^{'}
\left[
\frac{\epsilon_{{\bf
k}}-\mu+2n_0 V_0}{\hbar\Omega_{\bf k}}N(\hbar\Omega_{\bf k})\right.\nonumber\\
&\left.+\frac{\epsilon_{{\bf
k}}-\mu+2n_0 V_0-\hbar\Omega_{\bf k}}{2\hbar\Omega_{\bf k}}
\right],
\end{align}
and the disorder-induced depletion
\begin{align}
\label{eq:ndisorder}
\ndis=\frac{1}{V}{{\sum_{{\bf k}}}}^{'}n_0 \Rk
\frac{\left(\epsilon_{{\bf
k}}-\mu+n_0 V_0
\right)^2}{\hbar\Omega_{\bf k}^4}.
\end{align}
Here $N(\hbar\Omega_{\bf k})=(e^{\beta\hbar\Omega_{\bf k}}-1)^{-1}$
is the Bose distribution function.
The chemical potential can be eliminated from this expression by minimizing 
$\langle\Omega\rangle$,
given in Eq.~(\ref{eq:thermpotBog0}),
with respect to the condensate density $n_0$ [cf. Eq.~(\ref{eq:chempotBog0}) below]. In the mean-field approximation this yields
$\mu\simeq n_0V_0$, which corresponds to the Hugenholtz-Pines relation
for that approximation.
At this minimum, 
the thermodynamical potential of Eq.~(\ref{eq:thermpotBog0}) 
becomes (with $n_0\simeq \mu/V_0$)
\begin{align}
\label{eq:thermpotBogfin}
\langle\Omega\rangle\simeq
&
-\frac{1}{2}n_0^2 V_0 V+
{{\sum_{{\bf k}}}}^{'}\frac{1}{\beta}{\rm{ln}}\left(1-{\it e}^{-\beta\hbar\Omega_{{\bf
k}}}
\right)
\\
&
+\frac{1}{2}{{\sum_{{\bf k}}}}^{'}\left[\hbar\Omega_{\bf k}-\epsilon_{{\bf
k}}-n_0 V_0
\right]-
{{\sum_{{\bf k}}}}^{'}\frac{\Rk n_0}{\epsilon_{{\bf
k}}+2n_0 V_0},\nonumber
\end{align}
with
$\hbar\Omega_{\bf k}\simeq\sqrt{\epsilon_{\bf k}\left(\epsilon_{\bf k}+2n_0V_0
\right)}$.
Analogously, in the equation of state of Eq.~(\ref{eq:parnumBog}),
we get
\begin{align}
\label{eq:split1}
n'\simeq n_0\frac{8}{3}\left(\frac{n_0a^3}{\pi}\right)^{\frac{1}{2}}+
\frac{1}{V}{{\sum_{{\bf k}}}}^{'}
\frac{\epsilon_{{\bf
k}}+n_0 V_0}{\hbar\Omega_{\bf k}}N(\hbar\Omega_{\bf k}),
\end{align}
and 
\begin{align}
\label{eq:split2}
\ndis\simeq\frac{1}{V}{{\sum_{{\bf k}}}}^{'}
\frac{\Rk n_0}{\left(\epsilon_{\bf k}+2n_0V_0
\right)^2}.
\end{align}
Performing analytically the integration in Eq.~(\ref{eq:split2}), 
the number of condensate particles depleted by the disorder 
becomes \cite{Huang92}
\begin{align} 
\label{eq:disdepldelta}
\ndis=R_0 \frac{m^2}{8 ~\pi^{3/2}~\hbar^4}\sqrt{\frac{n_0}{a}}, 
\end{align} 
where we have made the replacement $V_0\rightarrow\rTtb$, that is, we have 
eliminated the unknown bare potential $V_0$ for the two-body scattering matrix
$\rTtb\equiv4\pi\hbar^2 a/m$ proportional to the $s-$wave scattering length $a$.
According to the limitations of the Bogoliubov approximation,
the theory is valid under the conditions $\ndis$,$n'\ll \nnu\simeq n $. 
The constraint $n'\ll n$ implies the diluteness condition $n^{1/3}a \ll 1$.
The condition $\ndis \ll n$
is equivalent to the inequality $R_0'\equiv m^2 R_0/8 \pi^{3/2}\hbar^4 
\left(n a\right)^{1/2}\ll 1$.
Associating to the strength $R_0$ of the disorder perturbation
a length scale 
defined as $d\equiv \left(2\pi\hbar^2/m\right)^2/R_0$, the condition on $\Rk '$ can be 
rewritten as $2\left(n^{1/3} a\right)^{1/2} d n^{1/3}/\sqrt{\pi}\gg 1$.
In the dilute limit $n^{1/3} a\ll 1$ this requires $d n^{1/3}\gg 1$,
{\it i.e.} the length scale
associated with the interaction energy due to the impurity potential
must be much larger than the interparticle distance $n^{-1/3}$.
The condition $R_0'\ll 1$ 
can also be reexpressed and elucidated 
by introducing the healing length of the condensate wave function. The latter is
defined as $\xi_{\rm heal}\equiv 1/\sqrt{8\pi n a}$. 
According to the theory of superfluidity  in BEC, the 
inverse of the healing length  
characterizes the upper boundary of the momenta in the phononic spectrum of the fluid.
At this wavelength, the energy of the excitations is of the order of $\Ek\simeq\mu$.
The condition $R_0'\ll 1$ is equivalent to 
$\sqrt{2}\pi \xi_{\rm heal}\ll d$. 
Therefore, the theory is valid 
when 
the energy of the excitations, induced by the impurity scattering, is far below the value
$\hbar/\xi_{\rm heal}^2$ that marks the crossover from the collective phononic excitations to the
single-particle excitations.

The lowest-order Hugenholtz-Pines condition $\mu=\nnu V_0$, neglects the effects
of quasi-particle interactions \cite{Beliaev58,Huang57} as well as the scattering
between the quasi-particles and the impurities \cite{Lopatin02}. Nevertheless, 
the beyond mean-field Beliaev corrections
to the leading order result $\mu=\nnu V_0$  
depend, 
both in the normal and in the disorder interactions,
only on two-body
collisions and can be calculated in the framework of the Bogoliubov theory \cite{PitaevskiiBook}.
Minimizing the thermodynamic potential of Eq.~(\ref{eq:thermpotBog0}) with respect to the 
condensate density, we have
\begin{align}
\label{eq:chempotBog0}
\mu= &n_0V_0+
\vnu\frac{1}{V}{{\sum_{{\bf k}}}}^{'}
\left[
\frac{2\epsilon_{{\bf
k}}-2\mu+3n_0 V_0}{\hbar\Omega_{\bf k}}N(\hbar\Omega_{\bf k})
\right.\nonumber\\
&\left.
+
\frac{2\epsilon_{{\bf
k}}-2\mu+3n_0 V_0-2\hbar\Omega_{\bf k}}{2\hbar\Omega_{\bf k}}
\right]\\
 &-\frac{1}{V}{{\sum_{{\bf k}}}}^{'}\Rk\left[
\frac{\epsilon_{{\bf
k}}-\mu+2n_0 V_0}{\hbar\Omega_{\bf k}^2}
\right.\nonumber\\
&\left.
-\frac{2n_0 V_0
\left(\epsilon_{{\bf k}}-\mu+n_0 V_0\right)
\left(2\epsilon_{{\bf k}}-2\mu+3n_0 V_0\right)
}{\hbar\Omega_{\bf k}^{4}}\right].\nonumber
\end{align}
In order to get the corrections to the mean-field result we 
substitute in the right-hand-side of equation Eq.~(\ref{eq:chempotBog0}) 
the zero-loop result $\mu=n_0V_0$ 
to obtain the next order correction to 
the relation between the chemical potential and the condensate density.
At $T=0$, the Bose distribution
$N(\hbar\Omega_{\bf k})$ can be neglected, and we have
\begin{align}
\mu&= n_0V_0+
2\vnu\frac{1}{V}{{\sum_{{\bf k}}}}^{'}
\frac{\epsilon_{{\bf
k}}+n_0 V_0-\hbar\Omega_{\bf k}}{2\hbar\Omega_{\bf k}}
\\&-n_0V_0^2\frac{1}{V}{{\sum_{{\bf k}}}}^{'}
\frac{1}{2\hbar\Omega_{\bf k}}
-\frac{1}{V}{{\sum_{{\bf k}}}}^{'}\Rk
\frac{\epsilon_{{\bf k}}-2n_0 V_0}{\left(\epsilon_{{\bf k}}+2n_0 V_0\right)^2}
\nonumber.
\end{align}
Subtracting the ultraviolet-divergent contribution 
$-n_0V_0^2(1/V){{\sum_{{\bf k}}}}^{'}1/2\epsilon_{{\bf k}}$ 
in the third term 
and eliminating in the above expression the bare potential $V_0$ for 
the two-body scattering matrix 
defined by the Lippmann-Schwinger equation $\rTtb=V_0-V_0
(1/V){{\sum_{{\bf k}}}}^{'}\frac{1}{2\epsilon_{{\bf k}}}\rTtb$, we find 
\begin{align}
\label{eq:serv}
\mu=&n_0\rTtb\left(1+ \frac{40}{3}\sqrt{\frac{n_0 {a}^3}{\pi}}\right)
\nonumber\\
&-\frac{1}{V}{{\sum_{{\bf k}}}}^{'}\Rk
\frac{\epsilon_{{\bf k}}}{\left(\epsilon_{{\bf k}}+2n_0 \rTtb\right)^2}+ 
\rTtb\ndis.
\end{align}
Using the equation for the number of 
particles in Eqs.~(\ref{eq:parnumBog}), (\ref{eq:split1}) and~(\ref{eq:split2}), we have that
 Eq.~(\ref{eq:serv}) can be rewritten as
\begin{align}
\label{eq:muLopTzero}
\mu=
n \rTtb\left(1+ \frac{32}{3}\sqrt{\frac{n {a}^3}{\pi}}\right)+\mur,
\end{align}
with
\begin{align}
\mur=-\frac{1}{V}{{\sum_{{\bf k}}}}^{'}\Rk
\frac{\epsilon_{{\bf k}}}{\left(\epsilon_{{\bf k}}+2n_0 \rTtb\right)^2}
\end{align}
in agreement with the zero-temperature result of Ref.~\cite{Lopatin02}.
Note that 
the  beyond mean-field correction
given by $\mur$ still  contains
an ultraviolet divergency,
which is not related to the interaction and was thus not yet removed by the renormalization
of the latter. Rather its origin lies in the fact that
we have considered a $\delta -$correlated  random potential.
In second order perturbation theory, it changes the original chemical potential
$\mu=\mu_{\rm bare}$
without the random potential, 
to the new value 
$\mu=\mu_{\rm bare}-\frac{1}{V^2}{{\sum_{{\bf k}}}}^{'}\frac{|U_{\bk}|^2}{\epsilon_{{\bf k}}}$.
Thus, after averaging, we must make the change $\mu=\mu_{\rm bare}\rightarrow
\mu+R_0(1/V){{\sum_{{\bf k}}}}^{'}1/\epsilon_{{\bf k}}$, which removes 
the divergency.
After renormalizing in this way, we have
\begin{align}
\label{eq:mur}
\mur=6~\rTtb\ndis,
\end{align}
which corresponds to a shift for the macroscopic compressibility $
\partial\left(\mur\right)/\partial n=\delta\chi/\chi_0=3 \ndis/n$ with respect
to the mean-field value $\chi_0=\rTtb n/m$ of the Bogoliubov theory
for a clean system.
Using the thermodynamic relation $\mu=\partial F/\partial N$, we can calculate 
the free energy $ F$ from Eq.~(\ref{eq:muLopTzero}).
At $T=0$ this coincides with the energy $E$ and 
we find \cite{Huang92,Lopatin02}
\begin{align}
\label{eq:thermpotBog3}
\frac{\langle E\rangle}{V}\!\simeq\!
\frac{2\pi a\hbar^2n^2}{m}
\left[1+\frac{128}{15}\left(\frac{n a^3}{\pi}\right)^{\frac{1}{2}}
+8 \Rk' 
\right].
\end{align}

\section{Superfluid component}
\label{sec:superfluidity}

In a Bose-Einstein condensate, random impurities 
constitute a source of incoherent scattering which tends to localize the 
condensate.
As shown by Huang and Meng  \cite{Huang92} 
the formation of 
local condensates in the minima of the random potential  reduces the superfluid component
of the fluid even at zero temperature, where, in the absence of disorder,
the whole fluid 
would be superfluid \cite{Kalatnikov62}. 
For the sake of completeness 
and in order to propose a simpler derivation,
we rederive in this section
the superfluid component
in the presence of weak disorder \cite{Huang92,Giorgini93,Lopatin02} 
by extending the Bogoliubov diagonalization method to a moving system.
The superfluid density $n_s$ is related to the total density by the relation
\begin{align}
\label{eq:supdensBog}
n_s=n-n_n,
\end{align}
where $n_n$ is the density of the normal component of the fluid.
In order to calculate $n_n$ , we consider the action of Eq.~(\ref{eq:actionmomspace})
when the gas is in motion. The 
moving reference system is related to the laboratory system 
by a Galilean transformation $\bx'=\bx+\velu t$ and $t'=t$.  
The fields in the moving system have a relative velocity $\velv$ with
respect to those in the inertial frame. They are related by the transformation
${\Psi^*}'\left(\bx',t'\right)={e}^{-i m\velv\bx/\hbar }\Psi^* \left(\bx,t\right)$   
and ${\Psi}'\left(\bx',t'\right)={e}^{i m\velv\bx/\hbar }\Psi \left(\bx,t\right)$.
Therefore,  in the new reference system, 
the action in Eq.~(\ref{eq:actionmomspace}) becomes
\begin{align}
\label{eq:actionmomspacemov}
S[a^*,a]=&\sum_{{\bf k},n}\left[-i\hbar\omega_n
+\hbar\bk\left(\velu-\velv\right)
+\epsilon_{{\bf k}}-\muef\right]
a^*_{{\bf k},n}a_{{\bf k},n}
\nonumber\\+&
\frac{1}{V}\sum_{{\bf
k,k',n}}a^*_{{\bf k},n}U_{{\bf k}-{\bf k}'}
a_{{\bf k}',n}\delta_{n,0}\\
+&\frac{1}{2}\frac{1}{\hbar\beta V}
\sum_{\stackrel{\bf
k,k',q}{n,n',m}} V_{\bf q}~a^*_{{\bf k+q},n+m}a^*_{{\bf k'-q},n'-m}
a_{{\bf k'},n'}a_{{\bf k},n},\nonumber
\end{align}
where the new chemical potential is defined as $\muef=\mu+m \velu\velv-m\velv^2/2$.
In the broken symmetry regime, the action in the new reference frame, 
up to the quadratic order in the fluctuations fields,
reads 
\begin{align}
\label{eq:quadractmov}
&S^{(2)}[a,a^*]=-\hbar\beta\muef n_0 V+\frac{1}{2}\hbar\beta n_0^2 V_0 V+
\hbar\beta n_0 U_{{\bf 0}}
\nonumber\\
+
&{{\sum_{{\bf k},n}}}^{'}\left[-i\hbar\omega_n
+\hbar\bk\left(\velu-\velv\right)
+\epsilon_{{\bf
k}}-\muef+2n_0 V_0\right]a^*_{{\bf k},n}a_{{\bf k},n}\nonumber \\
&+\left(
\frac{n_0}{V}\right)^{\frac{1}{2}}{\sum_{{\bf
k},n}}^{'}\left(a_{{\bf k},n}^*U_{{\bf k}}
+U_{-{\bf k}}a_{{\bf k},n}\right)\delta_{n,0}\nonumber\\
&+\frac{1}{2}n_0 V_0{\sum_{{\bf k},n}}^{'}(a^*_{{\bf k},n}a^*_{{\bf -k},-n}
+a_{{\bf k},n}a_{{\bf -k},-n}).
\end{align}
The latter action can again be diagonalized by the Bogoliubov transformation
described in Eqs.~(\ref{eq:Bogtrans})--(\ref{eq:bogospect}).
This is achieved by replacing $\mu\rightarrow\muef$
in the definition of $\uk$, $\vk$ and $\hbar\Omega_{\bf k}$, 
and by defining the new shift variable $z_{\pm{\bf k}}$ as
\begin{align}
\label{eq:zmovcond}
z_{\pm{\bf k}}=\left(n_0/V\right)^{\frac{1}{2}}
\frac{U_{\bf k}}{\hbar\Omega_{\bf k}\pm\hbar\bk\left(\velu-\velv\right)}
\left(u_{{\bf k}}-v_{{\bf k}}\right)^2.
\end{align}
Performing the diagonalization, the functional integration and the average 
over the disorder, we obtain the averaged thermodynamic potential
\begin{align}
\label{eq:thermpotBogmovcond}
\langle\Omega\rangle
&=-\muef n_0 V+\frac{1}{2}n_0^2 V_0 V
\nonumber\\&+
{{\sum_{{\bf k}}}}^{'}\frac{1}{\beta}\rm{ln}
\left\{1-{\it e}^{-\beta\left[\hbar\Omega_{{\bf
k}}
+\hbar\bk\left(\velu-\velv\right)\right]}
\right\}
\nonumber\\
&+\frac{1}{2}{{\sum_{{\bf k}}}}^{'}\left[\hbar\Omega_{\bf k}-\left(\epsilon_{{\bf
k}}-\muef+2n_0 V_0
\right)\right]
\nonumber\\
&-
{{\sum_{{\bf k}}}}^{'}n_0 \Rk
\frac{\left(\epsilon_{{\bf
k}}-\muef+n_0 V_0
\right)}{\hbar\Omega_{\bf k}^2-\left[\hbar\bk\left(\velu-\velv\right)\right]^2}.
\end{align}
Expanding for small $\hbar\bk\left(\velu-\velv\right)$ to second order, we have
\begin{align}
\label{eq:thermpotBogmovcondexp}
\langle\Omega\rangle
\simeq&-\muef n_0 V+\frac{1}{2}n_0^2 V_0 V+
{{\sum_{{\bf k}}}}^{'}\frac{1}{\beta}\rm{ln}
\left(1-{\it e}^{-\beta\hbar\Omega_{{\bf k}}}\right)
\nonumber\\
&+\frac{1}{2}{{\sum_{{\bf k}}}}^{'}\left[\hbar\Omega_{\bf k}-\left(\epsilon_{{\bf
k}}-\muef+2n_0 V_0
\right)\right]
\nonumber\\
&-
{{\sum_{{\bf k}}}}^{'}n_0 \Rk
\frac{\left(\epsilon_{{\bf
k}}-\muef+n_0 V_0
\right)}{\hbar\Omega_{\bf k}^2}\\
-{{\sum_{{\bf k}}}}^{'}& \frac{\hbar^2\bk^2}{3}
\left[\beta 
\frac{{\it e}^{-\beta\hbar\Omega_{{\bf k}}}}{2\left[
{\it e}^{-\beta\hbar\Omega_{{\bf k}}}-1\right]}
+\Rk n_0~\frac{2\epsk}{\Ek^4}
\right]\left(\velu-\velv\right)^2,\nonumber
\end{align}
where the term linear in $\velu-\velv$ vanishes as a consequence of the symmetry 
$\Ek=\Emenok$ of the Bogoliubov spectrum  of Eq.~(\ref{eq:bogospect}). 
Moreover, minimizing this thermodynamic potential with respect to $n_0$, 
we obtain the zero-loop result $\muef=n_0 V_0$
which gives $\hbar\Omega_{\bf k}\simeq\sqrt{\epsilon_{\bf k}\left(\epsilon_{\bf k}+2\muef
\right)}$. 
At this minimum, the momentum of the system can be calculated from the thermodynamic relation
\begin{align}
\label{eq:impulse}
\bp=-\frac{\partial\langle\Omega\left(T,V,\muef\left(\bu\right),\bu\right)
\rangle}{\partial\bu}{{\Big{|}}_{ T,V,\mu}}.
\end{align}
We find
\begin{align}
\label{eq:impulsecalc}
\bp=&m V n \velv
+
\frac{1}{3}{{\sum_{{\bf k}}}}^{'}\left[\beta\hbar^2\bk^2 
\frac{{\it e}^{-\beta\hbar\Omega_{{\bf k}}}}{\left[
{\it e}^{-\beta\hbar\Omega_{{\bf k}}}-1\right]}
\right.\\
&\left.
+\Rk ~n_0~ \hbar^2\bk^2\frac{4\epsk}{\Ek^4}
\right]\left(\velu-\velv\right)\nonumber,
\end{align}
where we have used the thermodynamical relation 
$n=-\left(1/V\right)\partial{\langle\Omega\rangle}/\partial{\mu}$
and the identity $\partial{\muef}/\partial{\bu}=m\velv$. Therefore, 
we can conclude that the density of the normal part of the fluid 
moving with velocity $\velu$ is given by \cite{Huang92}
\begin{align}
\label{eq:nordensBog}
n_n=&\frac{1}{V}{{\sum_{{\bf k}}}}^{'}\frac{2}{3}~\beta ~\epsilon_{\bf k}
\frac{e^{\beta\hbar
\Omega_{\bf k}}}{\left[e^{\beta\hbar\Omega_{\bf k}}-1\right]^2}+\frac{4}{3}\ndis.
\end{align}
Note that in two dimensions the derivation of the normal component due to the disorder 
is analogous, but
the factor $1/3$ in the formula for the thermodynamic potential in Eq.~(\ref{eq:nordensBog})
is replaced by a factor $1/2$. After the integration in two-dimensions, we get
\begin{align}
\label{eq:twoD}
n_n=\frac{1}{V}{{\sum_{{\bf k}}}}^{'}~\beta ~\epsilon_{\bf k}
\frac{e^{\beta\hbar
\Omega_{\bf k}}}{\left[e^{\beta\hbar\Omega_{\bf k}}-1\right]^2}+\frac{R_0 m^2}{8\pi^3\hbar^4 a},
\end{align}
in agreement with Refs. \cite{Meng,Giorgini93}.

The depletion of the superfluid density due to the disorder affects also the propagation of an external 
disturbance through the system, because the collective motion of the superfluid
component is ``hampered" by the component of the condensate localized in the minima
of the disorder potential. 
In order to see this effect, we now calculate  the corrections 
to the velocity of sound induced 
by the disorder at zero temperature. 
Let us assume that for weak disorder 
the dynamics of the superfluid component of the gas
can be described by the phenomenological two-fluid hydrodynamic equations \cite{Nozieres90}
\begin{align}
\label{eq:hydrodeq}
&\frac{\partial}{\partial t}n+{\bf{\nabla}}\left(\velv n_s+\veln n_n\right)=0\nonumber\\
&m \ddt \velv+\nabla\left(\mu+\frac{1}{2}m\velv^2\right)=0,
\end{align}
where $\mu$ is the chemical potential given in Eq.~(\ref{eq:muLopTzero}).
However, in that expression, we shall neglect the second-order Beliaev term due 
to the normal interactions in comparison to
the corrections due to the disorder, and we shall focus on the latter.
Then
Eqs.~(\ref{eq:hydrodeq}) 
represent a Landau ``two-fluid" model
for the superfluid condensate and the localized non-uniform condensate.
What is somewhat unusual is that, in this case, the normal component induced by the disorder
is at zero temperature and does not carry any entropy of the system.
Furthermore, we assume that for low frequency excitations, only
the superfluid component can react to the probe, while the localized normal
component remains stationary. Such a situation is familiar from 
the propagation of the fourth sound in $^4$He \cite{Atkins59} which is expressed
by the condition $\veln=0$.
If we restrict ourselves to the linear regime, we can write 
$n(t)=n+\delta n(t)$
and $\mu=\mu_0+\delta\mu$ with $\delta\mu=\left(\partial\mu/\partial n\right)\delta n$.
Then, Eqs.~(\ref{eq:hydrodeq}) give the equation of motion
\begin{align}
\label{eq:hydrodeq0}
&m \frac{\partial^2}{\partial t^2}\delta n-{\bf{\nabla}}\left[n_s\nabla\left(\frac{\partial\mu}{\partial n}\delta n\right)\right]=0.
\end{align}
From Eqs.~(\ref{eq:disdepldelta}) and~(\ref{eq:mur}) we have 
that $\left(\partial\mu/\partial n\right)=\rTtb\left(1+3\ndis/n\right)$. Moreover, 
from the result for the superfluid density in Eq.~(\ref{eq:supdensBog}) 
we have $n_s=n\left(1-4 \ndis/3n\right)$. Therefore,
Eq.~(\ref{eq:hydrodeq0}) can be put into the form
\begin{align}
\label{eq:hydrodeq1}
&\frac{\partial^2}{\partial t^2}\delta n-c^2{\nabla}^2\delta n=0,
\end{align}
which exhibits a phonon dispersion $\hbar\omega=c q$. 
Within this direct approach, 
the sound velocity
is found to be $c^2\simeq c_0^2\left(1+5\ndis/3 n\right)$ 
in agreement with Refs.
\cite{Giorgini93,Lopatin02}. Note that
$c_0^2=\rTtb n/m$ is the mean-field value of the Bogoliubov theory
for the clean system. The derivation of the sound mode we gave here is rather general 
and sufficiently simple to be generalized in order to calculate the effects of the disorder
on the frequencies of the collective modes in trapped gases \cite{Falco07}.

\section{Popov's theory}
\label{sec:popov}
At finite temperature the interactions of the thermal component of the gas
are described by
the contributions to the action beyond the quadratic order given   
in Eq.~(\ref{eq:quadract}).
In this section we extend the Bogoliubov approach of Section \ref{sec:Bog}
including these fluctuations according to the scheme of the 
Popov theory \cite{Popov85} which is designed for the temperature domain $k_B T>\mu$. 
In that approximation 
the cubic and quartic contributions are taken as
\begin{align}
\label{eq:cubicactHF}
S^{(3)}[a,a^*]\simeq\sqrt{\frac{n_0}{\hbar\beta V}}{\sum_{{\bf k},n}}^{'}{\tilde n}V_0
\left(a^*_{{\bf k},n}+a_{{\bf k},n}\right)
\end{align}
and
\begin{align}
\label{eq:quarticactHF}
S^{(4)}[a,a^*]\simeq
\frac{2}{\hbar\beta V}{\sum_{{\bf
k},{n}}}^{'}{\tilde n}V_0 a^*_{{\bf k},n}
a_{{\bf k},n},
\end{align}
where the temperature dependent total depletion ${\tilde n}\equiv\langle\langle
{\psi'}^{\star}\psi'\rangle_{\rm gr}\rangle$
is still to be determined.  
For our present thermodynamic considerations,
we  neglect the cubic terms following \cite{Popov85}
and include 
the quartic term. 
The cubic term in Eq.~(\ref{eq:cubicactHF}) only contributes in second and higher orders of $V_0$
and is taken into account by the introduction of the $\rTtb-$matrix below.
The new action is still diagonalized by the same Bogoliubov transformation 
Eqs.~(\ref{eq:Bogtrans})--(\ref{eq:bogospect})
but with the difference that the chemical potential $\mu$ has now to be replaced everywhere
by the new variable  
\begin{align}
\label{eq:modHP}
\mu'=\mu-2{\tilde n}V_0,
\end{align}
and that the condensate density $\nnu$ becomes strongly temperature
dependent.
Performing again the functional integral and the disorder average, the thermodynamic
potential of Eq.~(\ref{eq:thermpotBog0}) reads
\begin{align}
\label{eq:thermpotBog9}
\langle\Omega\rangle
=-&\mu n_0 V+\frac{1}{2}n_0^2 V_0 V+
{{\sum_{{\bf k}}}}^{'}\frac{1}{\beta}\rm{ln}\left(1-{\it e}^{-\beta\hbar\Omega_{{\bf
k}}
}
\right)
\nonumber\\
+&\frac{1}{2}{{\sum_{{\bf k}}}}^{'}\left[\hbar\Omega_{\bf k}-\left(\epsilon_{{\bf
k}}-\mu'+2n_0 V_0
\right)\right]\nonumber\\-&
{{\sum_{{\bf k}}}}^{'}\Rk n_0
\frac{\left(\epsilon_{{\bf
k}}
-\mu'+n_0 V_0
\right)}{\hbar\Omega_{\bf k}^2}.
\end{align}
Popov theory is equivalent to replacing 
in the contribution $n_0^2 V_0/2 $ to the pressure $\Omega/V$ 
the bare 
interaction $\vnu$ by the renormalized $\rTtb$-matrix 
and to adding the contributions $2 \nnu {\tilde n}\rTtb+ {\tilde n}^{2}\rTtb$
\cite{Popov85}. After these steps
the averaged thermodynamic potential can be rewritten as
\begin{align}
\label{eq:thermpotBog10}
\langle\Omega\rangle
=&-\mu n_0 V+\frac{1}{2}n_0^2 \rTtb V+2 \nnu {\tilde n}\rTtb+ {\tilde n}^{2}\rTtb
\nonumber\\&+
{{\sum_{{\bf k}}}}^{'}\frac{1}{\beta}\rm{ln}\left(1-{\it e}^{-\beta\hbar\Omega_{{\bf
k}}
}
\right)\nonumber
\\
&+\frac{1}{2}{{\sum_{{\bf k}}}}^{'}\left[\hbar\Omega_{\bf k}-\left(\epsilon_{{\bf
k}}-\mu'+2n_0 \rTtb
\right)\right]
\nonumber\\&-
{{\sum_{{\bf k}}}}^{'}\Rk n_0
\frac{\left(\epsilon_{{\bf
k}}-\mu'+n_0 \rTtb
\right)}{\hbar\Omega_{\bf k}^2}.
\end{align}
The equilibrium condition at fixed temperature is found by
minimizing the thermodynamic potential of Eq.~(\ref{eq:thermpotBog10})
with respect to $n_0$. 
Using the modified Hugenholtz-Pines relation $\mu'=n_0\rTtb$,
which fixes ${\tilde n}$ according to Eq.~(\ref{eq:modHP}) as ${\tilde n}=[({\mu}/{\rTtb})-n_0]/2$,
we obtain instead of Eq.~(\ref{eq:chempotBog0})
\begin{align}
\label{eq:muPopov0}
\mu'=& n_0\rTtb+n_0\rTtb\left(\frac{40}{3}\sqrt{\frac{n_0 {a}^3}{\pi}}\right)
+\nonumber\\
&+\rTtb\frac{1}{V}{{\sum_{{\bf k}}}}^{'}
\left[\frac{2\epsilon_{{\bf
k}}+n_0 \rTtb}{\hbar\Omega_{\bf k}}N(\hbar\Omega_{\bf k})-
N(\epsilon_{\bf k})\right]
\nonumber\\
&+2 \zeta\left(\frac{3}{2}\right)
\left(\frac{m k_B T}{2 \pi \hbar^2}\right)^{\frac{3}{2}}\rTtb
\nonumber\\
&-\frac{1}{V}{{\sum_{{\bf k}}}}^{'}\Rk
\frac{\epsilon_{{\bf k}}}{\left(\epsilon_{{\bf k}}+2n_0 \rTtb\right)^2}+\rTtb \ndis,
\end{align}
where $\ndis$ is given by the same expression as in Eq.~(\ref{eq:split2}).
In the ``high'' temperature limit
$k_B T\gg n_0\rTtb$, 
the main contribution to the momentum integral 
containing the Bose distribution
comes from the 
region $\epsilon_{\bf k}\leq n_0 \rTtb$
and the Bose distribution can be approximated as $N(x)\simeq k_B T/x$. 
Therefore, Eq.~(\ref{eq:muPopov0})
can be rewritten as
\begin{align}
\label{eq:muPopov1}
\mu'=
& n_0\rTtb\left(1+ \frac{40}{3}\sqrt{\frac{n_0 {a}^3}{\pi}}\right)
\nonumber\\
-3\rTtb&\left(\nnu\rTtb\right)^{\frac{1}{2}}\frac{m^{\frac{3}{2}}k_B T}{2 \pi\hbar^3}
+2 \zeta\left(\frac{3}{2}\right)
\left(\frac{m k_B T}{2 \pi \hbar^2}\right)^{\frac{3}{2}}\rTtb
\nonumber\\&-\frac{1}{V}{{\sum_{{\bf k}}}}^{'}\Rk
\frac{\epsilon_{{\bf k}}}{\left(\epsilon_{{\bf k}}+2n_0 \rTtb\right)^2}+ \rTtb\ndis.
\end{align}
In the limit of zero disorder $R_0=0$ this result reduces, of course, to that of Popov
\cite{Popov85}.
The new equation of state is
\begin{align}
\label{eq:parnumPop}
n=&-\frac{1}{V}\frac{\partial\langle\Omega\rangle}{\partial \mu}=n_0+
\frac{8}{3}\left(\frac{n_0a^3}{\pi}\right)^{\frac{1}{2}}
\nonumber\\&+
\frac{1}{V}{{\sum_{{\bf k}}}}^{'}
\frac{\epsilon_{{\bf
k}}+n_0 \rTtb}{\hbar\Omega_{\bf k}}N(\hbar\Omega_{\bf k})
\nonumber\\&+
\frac{1}{V}{{\sum_{{\bf k}}}}^{'}
\frac{\Rk n_0}{\left(\epsilon_{\bf k}+2n_0\rTtb
\right)^2}.
\end{align}
This latter equation has the same form as
the equation of state of the Bogoliubov theory 
as given in Eq.~(\ref{eq:parnumBog}) with (\ref{eq:split1}) and (\ref{eq:split2}),
but the domain of validity
and the details of the temperature dependence 
are, of course, different. 
In the ``high'' temperature region $k_B T\gg n_0\rTtb$ 
 where the Popov theory applies,
we can neglect the quantum depletion of the zero-temperature theory,
and the thermal depletion $n'=\left({1}/{V}\right){{\sum_{{\bf k}}}}^{'}
N(\hbar\Omega_{\bf k})\left[\left(
{\epsilon_{{\bf
k}}+n_0 \rTtb}\right)/{\hbar\Omega_{\bf k}}
\right]$
can be simplified as
\begin{align}
\label{eq:thermdepPop1}
n'\simeq-\left(\nnu\rTtb\right)^{\frac{1}{2}}
\frac{m^{\frac{3}{2}}k_B T}{2 \pi\hbar^3}
+ \zeta\left(\frac{3}{2}\right)
\left(\frac{m k_B T}{2 \pi \hbar^2}\right)^{\frac{3}{2}}.
\end{align}
Therefore, Eq.~(\ref{eq:parnumPop})
can be rewritten as
\begin{align}
\label{eq:parnumPop2}
n&=n_0+n\left(\frac{T}{\tcz}\right)^{\frac{3}{2}}
-\left(\nnu\rTtb\right)^{\frac{1}{2}}\frac{m^{\frac{3}{2}}k_B T}{2 \pi\hbar^3}
+\ndis.
\end{align}
The curve for the critical temperature $T_c$ in the Popov theory
can be obtained by putting $\nnu(T)$ equal to zero in Eq.~(\ref{eq:parnumPop2}). 
We note that the contribution in Eq.~(\ref{eq:parnumPop2}) due to the disorder
vanishes when approaching the critical point
because, according to Eq.~(\ref{eq:disdepldelta}), we have
$\ndis\propto \sqrt{\nnu(T)}$. Therefore,
we can argue that
even the presence of a random potential of the kind under consideration here, 
the Popov approximation  
does not shift the value of the critical
temperature 
away from that of an ideal Bose gas $\tcz$.
In the absence of disorder, the Popov approximation of a dilute Bose gas
describes a first-order phase transition: 
at the critical temperature $\tcz$ the condensate density $\nnu\left(T\right)$
exhibits a discontinuous jump to a finite value. The latter 
can be calculated analytically \cite{Shi98} 
as a function of the scattering length $a$. It can be shown, that the effect of the disorder
is to suppress this discontinuity
and that the jump vanishes for $R_0'$ larger
than some value ${\tilde R}_0'$.
 
Nevertheless, 
the transition to superfluidity occurs
at the temperature $T_s$ where the superfluid density vanishes,
which means, where the equation
\begin{align}
\label{eq:normdensPop}
n&=n_n=\frac{1}{V}{{\sum_{{\bf k}}}}^{'}\frac{2}{3}~\beta_{s} ~\epsilon_{\bf k}
\frac{e^{\beta_{s}\hbar
\Omega_{\bf k}}}{\left[e^{\beta_{s}\hbar\Omega_{\bf k}}-1\right]^2}
+\frac{4}{3}\ndis,
\end{align} 
is satisfied.
At ``high'' temperatures such that $k_B T_{s}\gg n_0\rTtb$, the latter equation
can be approximated as
\begin{align}
\label{eq:normdensPop1}
n
\simeq &
n\left(\frac{T_{s}}{\tcz}\right)^{\frac{3}{2}}
-\frac{2}{3}\left(\nnu\rTtb\right)^{\frac{1}{2}}
\frac{m^{\frac{3}{2}}k_B T_{s}}{2 \pi\hbar^3}
+\frac{4}{3}\ndis.
\end{align} 
Therefore, Eqs.~(\ref{eq:parnumPop2}) and~(\ref{eq:normdensPop1}) 
represent two coupled equation for $\nnu$
and $T_s$.
In the lowest order 
in the strength of disorder and interaction
we can approximate Eq.~(\ref{eq:parnumPop2}) 
with the ideal gas result
$n_0\simeq n
\left[1- \left(T/\tcz\right)^{\left(3/2\right)}
\right]$.
Substituting this in Eq.~(\ref{eq:normdensPop1}) and neglecting there the 
second term on the right-hand side, 
in the limit $R_0 ' \gg \left(n a^3\right)^{1/6}$,
we obtain an equation for the critical temperature $T_s$
as a function of the total density and the strengths
of the interaction and the disorder
\begin{align}
\label{eq:T_sPop}
\left[1- \left(T_s/\tcz\right)^{\frac{3}{2}}\right]^{\frac{1}{2}}
\simeq &\frac{4}{3} \sqrt{\frac{1}{n a}}\frac{R_0 m^2}{8 ~\pi^{3/2}~\hbar^4}\equiv
\frac{4}{3} R_0 '.
\end{align}
Solving for $T_s$ we find for the critical temperature where superfluidity disappears 
when coming from lower temperatures
\begin{align}
\label{eq:T_sPop1}
T_s/\tcz\simeq 1-\left(32/27\right){R_0 '}^2.
\end{align}
This result coincides with Eq.~(32) of Ref. \cite{Lopatin02}.
Here it has been derived 
assuming that near the critical temperature $T_s$ the 
interaction between the bosonic particles can be described by 
the temperature independent $\rTtb-$matrix.
This is consistent only if the critical temperature $T_s$ is not too close to the transition
temperature $\tcz$. The reason is that 
near the Bose-Einstein condensation point the 
quasi-particle interactions acquire a strong temperature 
dependence \cite{Fisher87,Bijlsma96,Shi98}.
As pointed out in Ref. \cite{Lopatin02}, the consistency condition is satisfied when 
$R_0 ' \gg \left(n a^3\right)^{1/6}$.
In the opposite regime, when $R_0 ' \ll \left(n a^3\right)^{1/6}$,
we will show in the next section that we have $T_s\simeq T_c$,
where $T_c$ is the Bose-Einstein condensation temperature including disorder effects
which depends linearly on $\Rk$. 
In that case, temperature effects on the particle-particle
scattering cannot be neglected and the two-body $\rTtb-$matrix 
of the Popov theory has to be replaced by the many-body
$\rTmbu-$matrix, which we shall do in the next section.

The failure of 
the Popov approximation in the regime
$R_0 ' \ll \left(n a^3\right)^{1/6}$
has a specific physical reason.
The Popov approximation, as considered 
here up to now, 
neglects the Hartree-Fock contribution
by which the presence of the impurities affects 
the scattering among two thermal particles. This can be seen, for example,
by calculating the relation between the chemical potential $\mu$ and the total density $n$
and comparing it with the result of Lopatin and Vinokur in Ref.~\cite{Lopatin02}.
Using Eqs.~(\ref{eq:muPopov1})--(\ref{eq:thermdepPop1}),
we find
\begin{align}
\label{eq:muPopfin}
\mu&=\rTtb n-\rTtb\frac{1}{\pi}\left(\mu'\right)^{\frac{1}{2}}
\frac{m^{\frac{3}{2}}k_B T}{\hbar^3}\\&+
\rTtb\zeta\left(\frac{3}{2}\right)
\left(\frac{m k_B T}{2 \pi \hbar^2}\right)^{\frac{3}{2}}
-\frac{1}{V}{{\sum_{{\bf k}}}}^{'}\Rk
\frac{\epsilon_{{\bf k}}}{\left(\epsilon_{{\bf k}}+2\mu'\right)^2}\nonumber.
\end{align}
We observe that in comparison with the theory of Ref.~\cite{Lopatin02}
we seem to miss in Eq.~(\ref{eq:muPopfin}) the Hartree-Fock term
$\rTtb R_0 m^3 k_B T/4 \pi^2 \hbar^6$.
A corresponding 
contribution seems also to be absent in the equation of state in Eq.~(\ref{eq:parnumPop2})
and in the normal density component of Eq.~(\ref{eq:normdensPop1}).
However, since we are in the regime $R_0 ' \gg \left(n a^3\right)^{1/6}$,
the latter contribution is in fact negligible and one obtains the value of $T_s$ as indicated
in Eq.~(\ref{eq:T_sPop1}). 
In order to access also the regime $R_0 ' \leq \left(n a^3\right)^{1/6}$,
in the next section,
we will extend the Popov theory to
the many-body $\rTmbu-$matrix approximation.
In this way we will find results similar to the one of Lopatin and Vinokur
but within
a gapless theory 
approaching the critical point from below.

\section{Many-body $\rT$-matrix}
\label{sec:many-bodyTmatrix}

In the vicinity of the critical temperature of Bose-Einstein condensation, the 
interactions between the quasi-particles 
are strongly renormalized by temperature effects
and vanish at the transition point for this reason \cite{Fisher87}.
For a clean system, Bijlsma and Stoof \cite{Bijlsma96} have shown that 
the quasi-particles interaction is well described by the  many-body ${\rTmbu}$-matrix
${\rTmb}$.
In the presence of weak disorder, the many-body ${\rTmbu}$-matrix
continues to obey, in lowest order, 
the same formal Bethe-Salpeter equation 
as in the absence of disorder
\cite{Bijlsma96}
\begin{align}
\label{eq:genmanybodTmatr}
&\rTmb({\bf k},{\bf k'},{\bf K};z)=V({\bf k}-{\bf k'})+\int \frac{d{\bf
k''}}{(2\pi)^3} V({\bf k}-{\bf k''})\nonumber \\
\times&\left\{\left[\frac{u_+^2u_-^2}{z-\hbar\Omega_+-\hbar\Omega_-}-
\frac{v_+^2v_-^2}{z+\hbar\Omega_++\hbar\Omega_-}\right]\right.(1+N_++N_-)\nonumber\\
+&\left[\left.\frac{u_-^2v_+^2}{z+\hbar\Omega_+-\hbar\Omega_-}-
\frac{u_+^2v_-^2}{z-\hbar\Omega_++\hbar\Omega_-}\right](N_+-N_-)\right\}
\nonumber\\ &
\,\,\,\,\,\,\,\,\,\,\,\,\,\,\,\,\,\,\,\,\,\,\,\,\,\,\,\,\,\,\,\,\,\,\,\,\,\,\,\,\,\,\,\,\,\,\,\,
\times\rTmb({\bf k''},{\bf k'},{\bf K};z),
\end{align}
where $N_+\equiv N(\hbar\Omega_+)$ and $N_-\equiv N(\hbar\Omega_-)$. The plus
sign denotes the momentum argument ${{\bf K}}/{2}+{\bf k''}$, and similarly
the minus sign denotes the argument ${{\bf K}}/{2}-{\bf k''}$.
In the low-temperature domain, 
where BEC-experiments with cold atoms are always realized,
the momentum and energy dependence
of the $\rTmb$-matrix can be neglected. This even applies to the domain
$k_BT_c\gg k_BT\gg\mu$
of sufficiently high temperatures
where the Popov approximation applies, to which we sometimes refer as the 
``high''-temperature domain
in this context.
Then we have
\begin{align}
\label{eq:BetheSalp2}
\rTmb&({\bf 0},{\bf 0},{\bf 0};0)=\vnu-\vnu T^{MB}({\bf 0},{\bf 0},{\bf
0};0)\\
&\times\int\frac{d{\bf k}}{(2\pi)^3}
\left[\frac{1}{2\hbar\Omega_{\bf k}}+\frac{\nnu V_0}{4\left(\hbar\Omega_{\bf k}\right)^3}\right]
\left[1+2N(\hbar\Omega_{\bf k})\right]\nonumber.
\end{align}
The integral on the right hand side contains an infrared divergency \cite{Nepo78}
caused by the well-known fact that the finite-temperature theory does not properly account
for the 
dynamics of the phase fluctuations in the infrared limit \cite{Andersen02}.
However, in the ``high'' temperature limit $na\lambda_{\rm th}^2\stackrel{<}{\sim} 1$
the infrared divergent term must, in fact, be dropped. 
This follows because at ``high'' temperatures the Bogoliubov spectrum
$\hbar\Omega_{\bf k}$ deviates from $\epsilon_{\bf k}-\mu'$ only for a very
small interval of momenta around zero, where $\epsilon_{\bf
k}\stackrel{<}{\sim}\nnu\rTtb<n \rTtb$. Therefore this phononic part of the spectrum
merely makes an asymptotically small contribution to all thermodynamic quantities. Therefore, 
we can set $u_{\bf k}=1$ and $v_{\bf k}=0$ in Eq.~(\ref{eq:genmanybodTmatr}).
With this, and after having eliminated the bare potential $\vnu$ 
by means of the Lippmann-Schwinger
equation for the $\rTtb-$matrix, we obtain
\begin{align}
\rTmb({\bf 0},{\bf 0},{\bf 0};0)&=\rTtb({\bf 0},{\bf 0};-2\mu')
- \rTtb({\bf 0},{\bf 0};-2\mu')
\nonumber\\\times\int&\frac{d{\bf k}}{(2\pi)^3}
\frac{N(\hbar\Omega_{\bf k})}{\hbar\Omega_{\bf k}}
\rTmb ({\bf 0},{\bf 0},{\bf 0};0).
\end{align}
Solving for $\rTmb({\bf 0},{\bf 0},{\bf 0};0)$ we find \cite{Bijlsma96,Shi98}
\begin{align}
\label{eq:manyb}
\rTmb({\bf 0},{\bf 0},{\bf 0};0)=\frac{\rTtb({\bf 0},{\bf 0};-2\mu')}{
1+\rTtb({\bf 0},{\bf 0};-2\mu')\int\frac{d{\bf k}}{(2\pi)^3}
\frac{N(\hbar\Omega_{\bf k})}{\hbar\Omega_{\bf k}}}.
\end{align}
The many-body $\rTmbu$-matrix approximation
in a dilute ultracold Bose gas can now be obtained
from the formulas of the
Popov theory as developed in the previous section just by replacing \cite{Shi98} there the
two-body $\rT$-matrix
$\rTtb({\bf 0},{\bf 0};-2\mu')\equiv\rTtb$ by the temperature dependent 
many-body $\rT$-matrix $\rTmb({\bf 0},{\bf 0},{\bf 0};0)$.
The excitation spectrum is thereby changed to
\begin{align}
\label{eq:exc}
\hbar\Omega_{\bf k}=\sqrt{\left[\epsilon_{\bf k}-\mu+\hbar\Sigma_{11}({\bf k},\omega_n)
\right]^2-\left[\hsi({\bf k},\omega_n)\right]^2},
\end{align}
where the self-energies are given by
\begin{align}
\label{eq:sigma12}
\hsi=\nnu\rTmb ({\bf 0},{\bf 0},{\bf 0};0)~,\nonumber\\
\hbar\Sigma_{11}=2nT^{MB}({\bf 0},{\bf 0},{\bf 0};0).
\end{align}
The condition $\mu'=\mu-\hbar\Sigma_{11}=\nnu\rTmb ({\bf 0},{\bf 0},{\bf 0};0)$ 
ensures the spectrum to be gapless.

The Bose factor in the many-body ${\rm T}-$matrix of Eq.~(\ref{eq:manyb})
leads to a 
temperature-dependent scattering length, defined as
\begin{align}
\label{eq:tempdepas}
\at\equiv m\rTmb\left({\bf 0},{\bf 0},{\bf 0};0\right)/4\pi\hbar^2.
\end{align}
For temperatures not too close to the Bose-Einstein critical temperature, 
where the usual
mean-field Popov theory of the previous section is valid, the denominator in Eq.~(\ref{eq:manyb}) 
represents a negligible correction. In that case, the many-body ${\rm T}-$matrix indeed
reduces to the two-body temperature-independent $\rTtb-$matrix  $\rTtb=4\pi\hbar^2 a/m$. 
However, approaching the critical region near the Bose-Einstein transition,  
the scattering between quasi-particles now becomes strongly temperature 
dependent \cite{Bijlsma96,Shi98}.
\begin{figure}[t]
\hspace{-10mm}
\subfigure[\label{fig:figAa}]
{\includegraphics[width=5.cm]{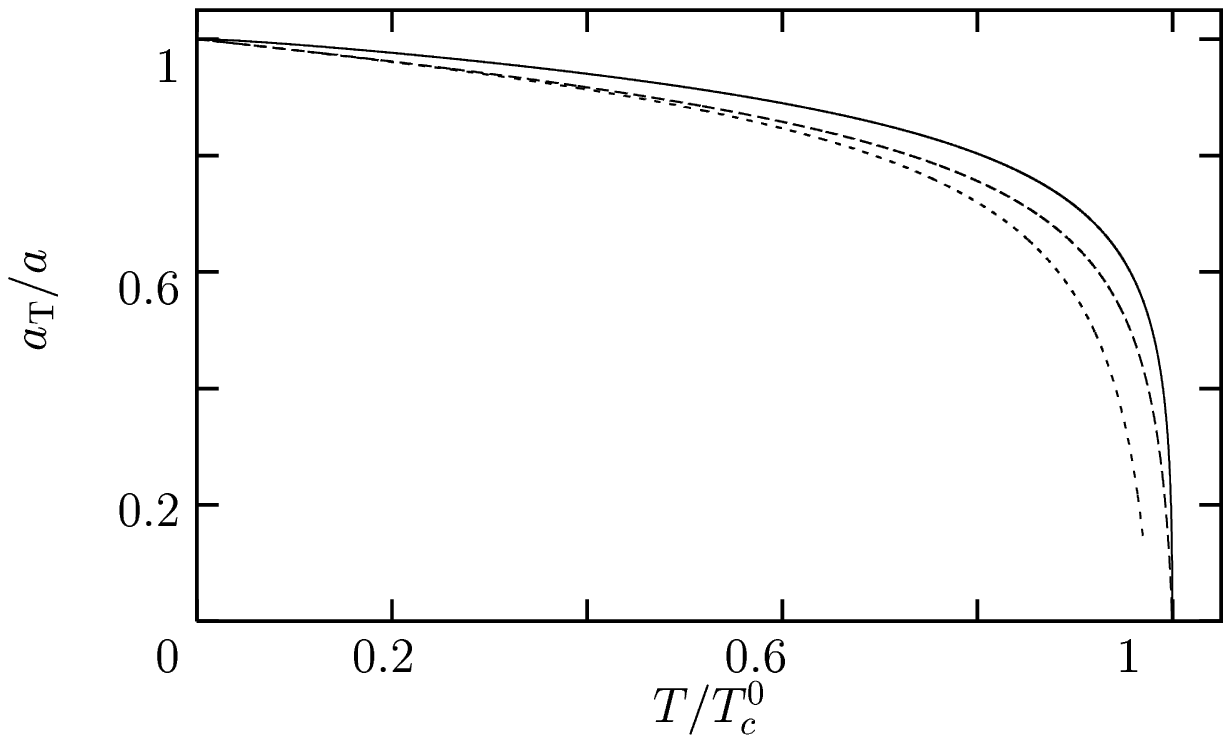}}\qquad
\subfigure[\label{fig:figAb}]
{\includegraphics[width=3.5cm]{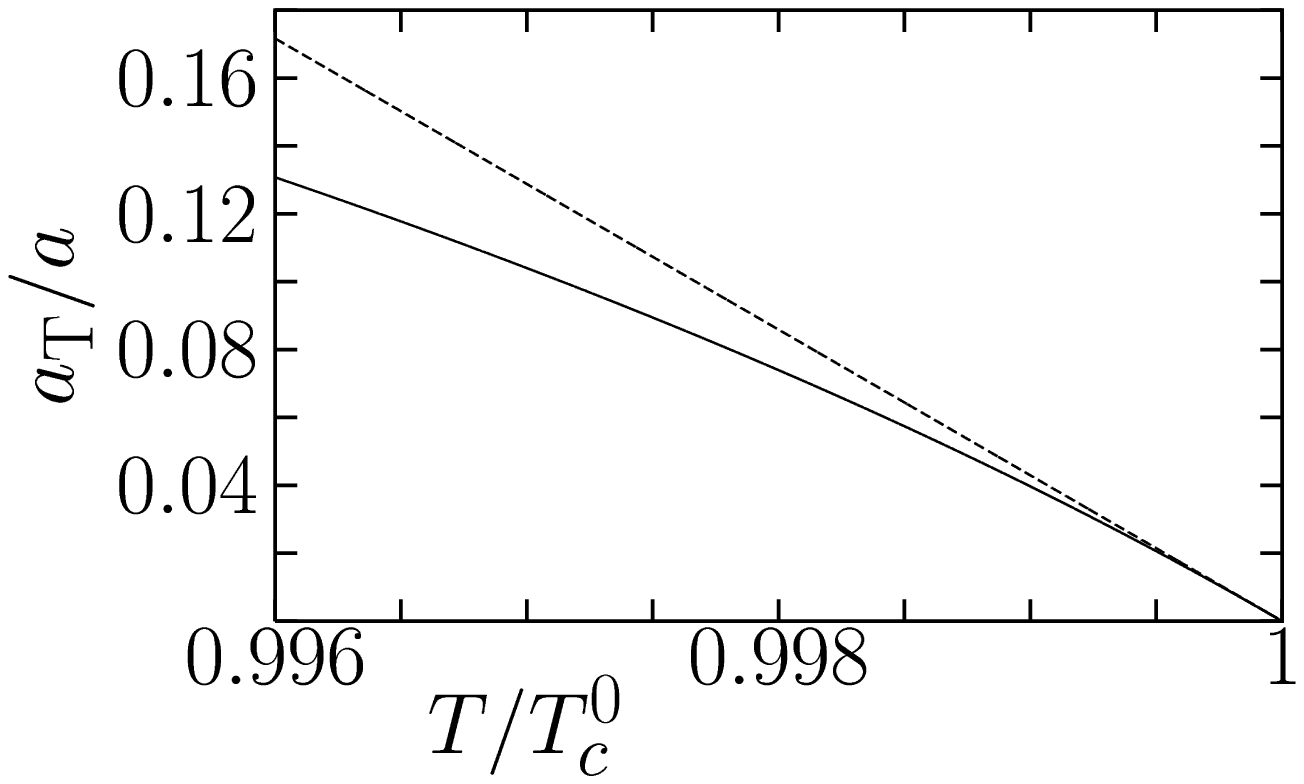}}
\caption{(a) Temperature dependent scattering
length as a function of the rescaled temperature $T/\tcz$
calculated numerically from Eq.~(\ref{eq:manyb}) (solid line)
and from the analytic expression of Eq.~(\ref{eq:manybanal}) (dashed line).
The dotted line shows the asymptotic limit described by Eq.~(\ref{eq:manybanalPopov})
valid in the Popov region for temperatures not too close to $\tcz$.
(b)  Temperature dependent scattering length as a function of the rescaled temperature $T/\tcz$
near $\tcz$. The lower line shows the numerical result 
from Eq~(\ref{eq:manyb}) compared to the asymptotic expression given
in Eq.~(\ref{eq:manybanalTmat}) described by the upper line. 
In both pictures the gas parameter of the Bogoliubov
theory has been chosen such as $(na^3)^{1/3}\sim 0.01$.   
 \label{fig:figuraA}}
\end{figure}
A more detailed insight can be gained  
by considering the ``high" temperature expansion of Eq.~(\ref{eq:manyb}) \cite{Shi98}. In that 
case, the latter reduces to a quadratic equation which can be solved analytically.
Using Eq.~(\ref{eq:tempdepas}) we find for the temperature dependent scattering length
\begin{align}
\label{eq:manybanal}
\at\simeq a\left[1+\frac{\left(\varrho\frac{T}{\tcz}\right)^2-
\left(\varrho\frac{T}{\tcz}\right)
\sqrt{\left(\varrho\frac{T}{\tcz}\right)^2+4\frac{n_0}{n}}}{2 \frac{n_0}{n}}\right].
\end{align}
where $\varrho\equiv\left[2\sqrt{\pi}/\zeta\left({3}/{2}\right)^{\frac{2}{3}}\right]
\left(n a^3\right)^{1/6}$. In \figurename~\ref{fig:figuraA} the curve resulting from
the``high" temperature expansion of  
Eq.~(\ref{eq:manybanal}) is compared with the curve obtained from 
Eqs.~(\ref{eq:manyb}) and~(\ref{eq:tempdepas}).
For temperatures not too close to the critical point, 
we have $\left(\varrho{T}/{\tcz}\right)^2\ll 4({n_0}/{n})$
and the curve of Eq.~(\ref{eq:manybanal})
can be approximated by
\begin{align}
\label{eq:manybanalPopov}
\at\simeq a\left[1-2\frac{\sqrt{\pi}}{\zeta\left({3}/{2}\right)^{\frac{2}{3}}}\left(n a^3\right)^{1/6}
\frac{T}{\tcz}\frac{1}{\sqrt{n_0/n}}\right],
\end{align}
while 
for temperatures just below the critical temperature $\tcz$
the scattering length between quasi-particles 
becomes
a universal function of the density and temperature
given by  
\begin{align}
\label{eq:manybanalTmat}
\at\simeq \frac{\zeta\left({3}/{2}\right)^{\frac{4}{3}}}{4\pi}\frac{n_0}{n^{4/3}},
\end{align}
which vanishes 
at the critical temperature $\tcz$ \cite{Fisher87,Bijlsma96}
with the same power as the condensate density.
The two different asymptotic regimes of the expression in Eq.~(\ref{eq:manybanal})
are also shown in \figurename~\ref{fig:figuraA}.
Note that \figurename~\ref{fig:figAa} 
includes also the zero-temperature limit where the Popov approximation
as well as the many-body ${\rm T}-$matrix are infrared divergent.
However, this is an asymptotic small region because
the condition of validity of the Popov approximation
defined by $n a \lambda_{\rm th}^2\stackrel{<}{\sim}1$ is equivalent to impose 
the condition $T/\tcz \stackrel{>}{\sim}
[\zeta\left({3}/{2}\right)^{\frac{2}{3}}/4\pi](na^3)^{1/3}$.

\begin{figure}[t]
\hspace{-1.5cm}
\includegraphics[width=8.6cm]{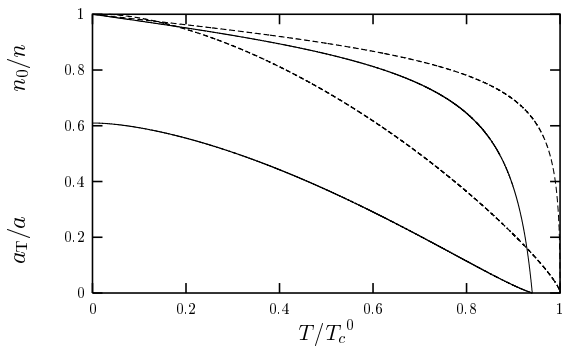} 
\caption{The lower and the upper
dashed lines describe the
condensate density $n_0$  and the scattering length $\at$ obtained
by solving self-consistently Eqs.~(\ref{eq:manybanal})
and~(\ref{eq:parnumTmanybod1}) in absence of disorder, {\it i.e.}, when $R=0$. 
In this case, both quantities vanish at $T=\tcz$ and no shift of the critical temperature
is induced \cite{Bijlsma96,Shi98}. 
The gas parameter of the Bogoliubov theory has been chosen 
the same as in \figurename~\ref{fig:figuraA}.
The solid lines show the solution for finite disorder.
We see that the temperature at which they vanish is shifted with respect
to the critical temperature of the ideal gas. In order to make clearly visible
the effect we have considered in the figure the case $R'_0=0.5$.
However, for such a value
the condensate depletion due to the disorder at zero temperature is already $\sim 0.4\%$ 
of the total density. Note also that, in contrast with Popov theory, 
for each value of $R'_0$ the curve for $n_0$ obtained from the 
many-body ${\rm T}-$matrix approximation exhibits a second-order phase transition.  
}
\label{fig:figuraC}
\end{figure}

By applying these results to the 
equation of state in Eq.~(\ref{eq:parnumPop2}) the latter can be rewritten as
\begin{align}
\label{eq:parnumTmanybod1}
n=n_0+&n\left(\frac{T}{\tcz}\right)^{\frac{3}{2}}
-\left[n_0
\frac{4\pi\hbar^2 \at}{m}
\right]^{\frac{1}{2}}\frac{m^{\frac{3}{2}}k_B T}{2 \pi\hbar^3}
\nonumber\\&
+R_0 \sqrt{\frac{n_0}{\at}}\frac{m^2}{8 ~\pi^{3/2}~\hbar^4}.
\end{align}
For a clean system neither the two-body nor the many-body 
$\rTmbu$-matrix theory induce 
a shift of the critical temperature
\cite{Bijlsma96}. In contrast to this, we find 
that  
in the presence of delta correlated disorder the effects described by the many-body
$\rTmbu$-matrix induce such a shift of $T_c$. This is illustrated in \figurename~\ref{fig:figuraC},
where the solutions of the coupled equations~(\ref{eq:manybanal})
and~(\ref{eq:parnumTmanybod1})
for $\at$ and the condensate density $n_0$ 
are shown as function of the temperature
in absence and in presence of disorder.
The origin of the shift
becomes evident when
considering the limit $T\rightarrow \tcz$ in Eq.~(\ref{eq:parnumTmanybod1}). 
If we insert in that equation for
the temperature dependence of the condensate density that of an ideal Bose gas,
the term due to the disorder does not vanish at $\tcz$ as it would in the simple Popov theory.
This follows because
the temperature dependent scattering length $\at$ goes to zero with the same exponent
as the condensate density $n_0$, and both effects cancel.
Thus we get 
\begin{align}
\label{eq:parnumTmanybodatTc}
n&=n\left({T}/{\tcz}\right)^{\frac{3}{2}}
+R_0 {m^3 k_B \tcz}/{8 \pi^2\hbar^6},
\end{align}
which is solved by
\begin{align}
\label{eq:Tcshift}
T_c\simeq 
\tcz\left(1-\eta\right),
\end{align}
where $\eta\equiv R_0{m^3 k_B \tcz}/{12 \pi^2\hbar^6 n}
=2\pi/[3 \zeta\left(3/2\right)^{2/3}d n^{1/3}]\ll 1$.
This value differs from the Hartree-Fock result found 
by Lopatin and Vinokur in Ref.~\cite{Lopatin02} by a factor of 1/2.
Note, however, that in contrast to Ref.~\cite{Lopatin02}, 
our gapless $\rTmb$-matrix theory 
approaches the critical point from below.
In fact, the theoretical descriptions on the two sides of the critical points
are remarkably different as a consequence of the difference between 
the two phases on both sides.
The result of Lopatin and Vinokur has been recently confirmed 
by Zobay in Ref.~\cite{Zobay06} by means of a 
one-loop Wilson renormalization group 
calculation. This latter method approaches the critical point from above  
as well but
takes into account 
critical fluctuations which are non-perturbative in the interaction.
The numerical solution of the renormalization group equations 
shows that the critical fluctuations lead only to small
corrections with respect to 
the result of Ref.~\cite{Lopatin02}.

\begin{figure}[t]
\vspace{1cm}
\hspace{-1.5cm}
\includegraphics[width=7.6cm]{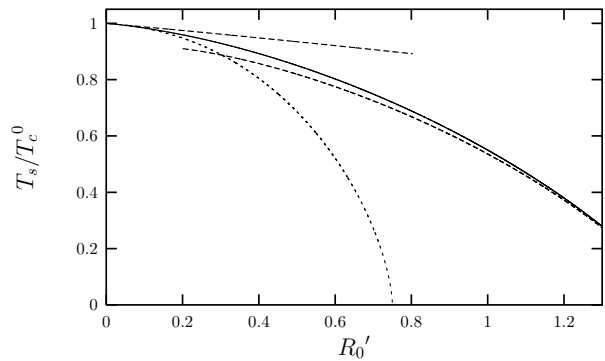} 
\caption{ The solid line indicates the critical temperature $T_s$ of the normal to superfluid transition as a function of the
dimensionless disorder parameter $R'_0$ obtained by solving self-consistently
Eq.~(\ref{eq:manybanal}), Eq.~(\ref{eq:parnumTmanybod1}),
and~Eq.~(\ref{eq:normdensTmanybody1}) for $(na^3)^{1/3}= 0.01$.
The upper dashed line is obtained by approximating 
Eq.~(\ref{eq:manybanal}) with the asymptotic solution of Eq.~(\ref{eq:manybanalTmat})
which is valid when $T_s\simeq T_c$ for $R_0 ' \leq {\left(n a^3\right)^{{1}/{6}}}$. 
In contrast, the lower dashed line describes the regime  $R_0 ' \geq {\left(n a^3\right)^{{1}/{6}}}$
where  $T_s\ll T_c$ and Eq.~(\ref{eq:manybanal}) can be approximated with 
Eq.~(\ref{eq:manybanalPopov}).  
The comparison between these three curves describes clearly 
how the temperature effects in the many-body $\rTmb -$matrix
determine
the crossover from the 
quadratic regime to the linear regime 
for the
dependence of $T_s/\tcz$ on the disorder coupling constant $\Rk'$.  
In addition, the short-dashed line shows the effects of neglecting the self-consistency in the equation 
of state~(\ref{eq:parnumTmanybod1}), 
where
the condensate density is approximated
with the ideal gas result $n_0=n[1-(T_s/\tcz)^{3/2}]$.
For $R_0 ' \geq {\left(n a^3\right)^{{1}/{6}}}$ the short-dashed line coincides with the analytical result
of the Popov theory given by Eq.~(\ref{eq:T_sPop1}). 
}
\label{fig:figuraD}
\end{figure}

The many-body $\rTmbu-$matrix  approximation is also
useful to understand 
the dependence on the disorder of
the superfluid transition $T_s$.
Equation~(\ref{eq:normdensPop1})  
can be rewritten as
\begin{align}
\label{eq:normdensTmanybody1}
n
\simeq 
n\left(\frac{T_s}{\tcz}\right)^{\frac{3}{2}}&
-\frac{2}{3}\left[{n_0}
\rTmb\right]^{\frac{1}{2}}\frac{m^{\frac{3}{2}}k_B T_s}{2 \pi\hbar^3}
\nonumber\\&
+\frac{4}{3}
R_0 \sqrt{\frac{n_0}{\at}}\frac{m^2}{8 ~\pi^{3/2}~\hbar^4}.
\end{align} 
Eq.~(\ref{eq:normdensTmanybody1}) together with Eqs.~(\ref{eq:manybanal})
and~(\ref{eq:parnumTmanybod1}) evaluated at $T=T_s$ constitute a set of closed 
equations for $T_s$, ${n_0}_{|T=T_s}$ and ${\at}_{|T=T_s}$.
The self-consistent solution of the three coupled equations is shown in 
\figurename~\ref{fig:figuraD}. 
When $R_0 ' \gg {\left(n a^3\right)^{{1}/{6}}}$  we have that $T_s$ lies well below $\tcz$
in the domain, where the many-body $\rTmbu-$matrix is given essentially 
by the temperature independent 
$\rTtb$-matrix. 
In that limit $T_s$ is given by Eq.~(\ref{eq:T_sPop1})
of the Popov theory.
However, when $R_0 ' \leq {\left(n a^3\right)^{{1}/{6}}}$, one expects 
$T_s \stackrel{<}{\sim} \tcz$
and the temperature effects on the quasi-particles scattering become important.
In that case, we can use the expansion 
of the many-body $\rTmbu-$matrix near the critical point 
given in Eq.~(\ref{eq:manybanalTmat}). 
From  Eq.~(\ref{eq:normdensTmanybody1}) to the lowest order in the disorder
and in the normal interaction this leads to
$T_s
\simeq \tcz\left(1-4\eta/3\right)$,
which is smaller than 
the critical temperature $T_c$ given in Eq.~(\ref{eq:Tcshift}) due to the factor $4/3$.

Therefore, we find that the presence of 
disorder shifts
the critical temperature for the superfluid transition
below the line of the Bose-Einstein condensation temperature. 
The shift increases monotonically
as a function of the disorder strength $\Rk$. 
For $R_0 ' \leq {\left(n a^3\right)^{{1}/{6}}}$ the shift 
is linear in $\Rk$.
At larger values of the disorder strength, where
$R_0 ' \gg {\left(n a^3\right)^{{1}/{6}}}$, it becomes quadratic in $\Rk$.
The crossover between the two regimes
is described accurately in \figurename~\ref{fig:figuraD}. 
Note that, in that figure, the results obtained 
in the region $\Rk'\sim 1$
have to be considered as an extrapo\-lation to the 
more complicate regime of strong disorder \cite{Navez06,Yukalov06}.

Moreover, the superfluid density can be calculated from 
the relation $n_s=n-n_n$ with the normal density given by the right-hand side 
of Eq.~(\ref{eq:normdensTmanybody1}) evaluated at temperatures $T<T_s$.
In contrast with the Bogoliubov approach 
of Ref.~\cite{Huang92}, 
our theory includes important finite-temperature correlations between quasi-particles 
and
we find that the superfluid density decreases monotonically as function of the temperature.
This result 
is in agreement with the diagrammatic theory based on the replica method developed in 
Ref.~\cite{Lopatin02}.

Here, we have not discussed the damping of the sound due to the impurity 
scattering. 
This damping has so far been studied only at zero temperature 
in Refs.~\cite{Giorgini93,Lopatin02}.
The many-body $\rTmbu -$matrix theory developed here,
neglects the finite lifetime of the quasi-particles and cannot describe 
damping phenomena for this reason.

\section{Conclusions and outlook}
\label{sec:conc}
In conclusion, we have extended in this paper the
perturbative approach developed by Huang and Meng in Ref.~\cite{Huang92}
for a homogeneous superfluid dilute Bose gas in the presence
of weak disorder. We have shown that such an extension to finite temperatures
is achieved via a suitable combination of the mean-field Popov approximation and  the 
many-body $\rTmbu -$matrix approximations. 
In particular this allows to make contact with results of
second-order perturbation theory at finite temperature
developed in Ref.~\cite{Lopatin02} by means of the replica method.
E.g. the shifts of the two different critical temperatures for the appearance of a condensate
density
and of a superfluid density have here been calculated when coming from the low-temperature
side. 

The theory could have other interesting applications.
Using the notion of the quasi-condensate \cite{Popov85},
Andersen {\it et al}. \cite{Andersen02} have shown that 
the many-body $\rTmbu -$matrix theory
can describe the 
thermodynamics of a clean two-dimensional Bose gas
near the 
Kosterlitz-Thouless 
superfluid transition, and it seems plausible that this continues to be true also for
weak disorder.
Therefore, the approach presented here 
could be useful in order to
study the long-standing problem 
concerning the disorder-induced shift on the superfluid transition 
temperature in a two-dimensional Bose gas.


\begin{acknowledgments}
We acknowledge financial support from 
the German Research Foundation [DFG] through
SFB/TR 12 ``Symmetries and Universality in Mesoscopic Systems".
\end{acknowledgments}

\end{document}